\documentclass[12pt]{iopart}

\usepackage{iopams}
\usepackage{graphicx}

\begin{document}

\title[Quantum filter reduction for measurement-feedback control]{Quantum filter reduction for measurement-feedback control via unsupervised manifold learning}

\author{Anne E B Nielsen$^{1,2}$, Asa S Hopkins$^3$ and Hideo Mabuchi$^2$}

\address{$^1$ Lundbeck Foundation Theoretical Center for Quantum
System Research, Department of Physics and Astronomy,
Aarhus University, 8000 \AA rhus C, DK}
\address{$^2$ Edward L. Ginzton Laboratory, Stanford University, Stanford CA 94305, USA}
\address{$^3$ Physical Measurement and Control 266-33, California Institute of Technology, Pasadena CA 91125, USA}
\eads{\mailto{ebnielse@stanford.edu}, \mailto{asa@caltech.edu}, \mailto{hmabuchi@stanford.edu}}

\begin{abstract}
We derive simple models for the dynamics of a single atom coupled to a cavity field mode in the absorptive bistable parameter regime by projecting the time evolution of the state of the system onto a suitably chosen nonlinear low-dimensional manifold, which is found by use of local tangent space alignment. The output field from the cavity is detected with a homodyne detector allowing observation of quantum jumps of the system between states with different average numbers of photons in the cavity. We find that the models, which are significantly faster to integrate numerically than the full stochastic master equation, largely reproduce the dynamics of the system, and we demonstrate that they are sufficiently accurate to facilitate feedback control of the state of the system based on the predictions of the models alone.
\end{abstract}

\maketitle

\section{Introduction}\label{1}

A two-level atom coupled to a driven and observed cavity mode exhibits a variety of interesting dynamics \cite{armen}, including scenarios where trajectories of the atom-cavity state tend to localize transiently but jump between multiple regions of phase space on longer timescales. Here and in related work, we loosely refer to such behaviour as 'bistability', and we use the term 'stable region' or 'attractor' to refer to the local regions of phase space, where the system spends most of its time. Bistable systems have potential applications as memory units and switches, and this motivates studies of ways to understand and control their dynamics. The phase bistable regime, where the system has two stable regions with different values of the phase of the cavity field, has been investigated in several papers \cite{alsing,kilin,wiseman,handel}, and it has been demonstrated that quantum jumps between the two stable regions can be observed in the photo current from a homodyne detector monitoring the field leaking out of the cavity \cite{wiseman}. As shown in \cite{savage}, there is also an absorptive bistable regime, for which the stable regions have different values of the amplitude of the cavity field mode. Quantum jump behaviour is observed in this case as well, but it is more complicated to obtain simple approximate descriptions of the dynamics due to lack of symmetry between the two stable regions \cite{hopkins}, and we thus consider this regime in the following. Examples of experimental investigations of bistability in cavity quantum electrodynamic systems are provided in \cite{grant,rempe,chapman,gupta}.

The time evolution of the state of a continuously monitored quantum system is governed by a stochastic differential equation, but it is typically a very slow process to integrate this equation numerically due to the large dimensionality of the Hilbert space. This is, in particular, a problem, if we would like to control the system dynamics through feedback, since, in that case, it is necessary to track the state of the system in real time. In many cases, it turns out that the dynamics does not explore all degrees of freedom in the full Hilbert space, and this opens the way to develop simple low-dimensional models, which can, at least approximately, predict the time evolution of the state of the system. One way to obtain such models is to project the system dynamics onto an affine linear subspace of low dimension, and this technique has turned out to be very successful in the case of phase bistability \cite{handel,hopkins}, while the results for absorptive bistability are less satisfactory \cite{hopkins}. It is, however, quite possible that improved results can be obtained by considering the more flexible case of projection onto a nonlinear manifold. In the present paper, we demonstrate that the latter approach provides simple models of absorptive bistable dynamics, which are sufficiently accurate to allow feedback control of the state of the system.

The paper is structured as follows. In section \ref{2}, we introduce the system and integrate the stochastic master equation to provide examples of absorptive bistable dynamics and quantum jumps. In section \ref{3} we derive reduced models of the behaviour of the system by first identifying a low-dimensional manifold, which captures most of the dynamics, and then projecting the full system dynamics onto that manifold. The ability of the reduced model to reproduce the results of the full model is investigated in section \ref{4}. Finally, in section \ref{5}, we demonstrate that a feedback scheme, which builds only on predictions of a reduced model, can be used to hold the system at one of the stable regions. Section \ref{6} concludes the paper.

\section{Absorptive bistability and quantum jumps}\label{2}

\begin{figure}
\center
\includegraphics[clip,width=0.5\textwidth]{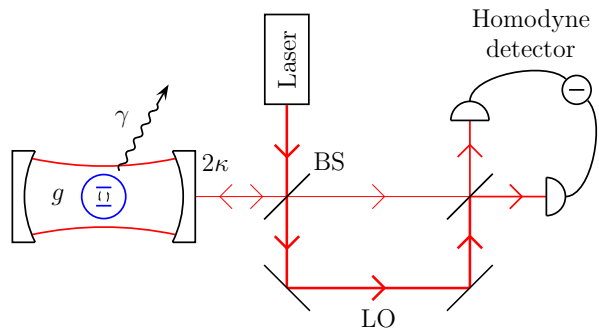}
\caption{Two-level atom in a cavity probed with a coherent laser beam. BS is a beam splitter of low reflectivity, and LO is the local oscillator.}\label{figure1}
\end{figure}

Our model system is a two-level atom with ground state $|g\rangle$ and excited state $|e\rangle$ coupled to a cavity field mode with coupling strength $g$ as illustrated in figure~\ref{figure1}. The cavity mode, which decays at a rate $2\kappa$, is driven by a coherent laser beam, and light reflected from the cavity is observed with a homodyne detector. The excited state of the atom decays at a rate $\gamma$ by spontaneous emission, but since the emitted photons travel in random directions, it is so far not experimentally feasible to detect all of them with high efficiency. We thus assume no detection of spontaneously emitted photons in the following and use a density operator $\rho$ to represent the state of the atom and the cavity field mode.

The time evolution of $\rho$ in a frame rotating with the frequency of the drive laser is determined by the stochastic master equation (see, for instance, \cite{nm6} for a derivation)
\begin{eqnarray}\label{SME}
\fl\rmd\rho=-\frac{\rmi}{\hbar}[H,\rho]\rmd t
+\kappa(2\hat{a}\rho\hat{a}^\dag-\hat{a}^\dag\hat{a}\rho
-\rho\hat{a}^\dag\hat{a})\rmd t
+\frac{\gamma}{2}(2\sigma\rho\sigma^\dag-\sigma^\dag\sigma\rho
-\rho\sigma^\dag\sigma)\rmd t\nonumber\\
+\sqrt{2\kappa}\left\{\rho\hat{a}^\dag \rme^{\rmi\phi}+\hat{a}\rho\rme^{-\rmi\phi}
-\Tr\left[\left(\hat{a}^\dag\rme^{\rmi\phi}+
\hat{a}\rme^{-\rmi\phi}\right)\rho\right]\rho\right\}\rmd W
\end{eqnarray}
with Hamiltonian
\begin{equation}
H=\hbar\Delta_{\rm c}\hat{a}^\dag\hat{a}+\hbar\Delta_{\rm a}\sigma^\dag\sigma +\rmi\hbar g_0(\hat{a}^\dag\sigma-\hat{a}\sigma^\dag)
+\rmi\hbar\mathcal{E}(\hat{a}^\dag-\hat{a}).
\end{equation}
Here, $\hat{a}$ is the cavity field annihilation operator, $\sigma=|g\rangle\langle e|$ is the atomic lowering operator, $\Delta_{\rm c}$ is the detuning between the cavity resonance frequency and the frequency of the drive laser, $\Delta_{\rm a}$ is the detuning between the atomic transition frequency and the frequency of the drive laser, and $\mathcal{E}=\sqrt{2\kappa}\beta$, where $|\beta|^2$ is the average number of photons in the probe beam arriving at the cavity input mirror per unit time and $\beta$ is assumed to be real. The phase $\phi$, which is varied experimentally by varying the relative phase of the probe beam and the local oscillator, determines which quadrature of the cavity field is detected. The $x$-quadrature is measured for $\phi=0$ and the $p$-quadrature is measured for $\phi=\pi/2$. Finally, the Wiener increment $\rmd W$ is a Gaussian stochastic variable with mean $0$ and variance $\rmd t$, which is related to $\rmd y$, the observed homodyne photo current (in units of photons per time) integrated from $t$ to $t+\rmd t$ divided by the square root of the average number of photons per unit time in the local oscillator beam, through
\begin{equation}\label{dy}
\rmd y=\rmd W+\sqrt{2\kappa}\Tr(\hat{a}\rho\rme^{-\rmi\phi}
+\rho\hat{a}^\dag\rme^{\rmi\phi})\rmd t.
\end{equation}
In an experiment, the photo current is measured as a function of time, and we can eliminate $\rmd W$ between (\ref{SME}) and (\ref{dy}). In numerical simulations, on the other hand, we use a random number generator to obtain realizations of $\rmd W$ and integrate (\ref{SME}) directly.

Useful insight into the dynamics predicted by (\ref{SME}) can be obtained through various semiclassical approximations \cite{armen,mabuchi}. We shall not pursue such models further here, but simply choose a set of parameters for which the dynamics of the system has been shown to be absorptive bistable \cite{hopkins}: $\Delta_{\rm c}/\gamma_\bot=0$, $\Delta_{\rm a}/\gamma_\bot=0$, $\kappa/\gamma_\bot=0.1$, $g_0/\gamma_\bot=\sqrt{2}$ and $\mathcal{E}/\gamma_\bot=0.56$, where $\gamma_\bot=\gamma/2$ is the transverse atomic decay rate. We assume throughout that the initial state of the system is the state with zero photons in the cavity and the atom in the ground state, and we use the second order derivative free predictor-corrector method of \cite{kloeden} to integrate (\ref{SME}). The expectation value of the number of photons in the cavity is typically below 22, and we thus truncate the basis of the Hilbert space of the cavity field at 59 photons, leading to a density matrix of dimension $120\times120$.

\begin{figure}
\flushright
\includegraphics[clip,width=0.41\textwidth]{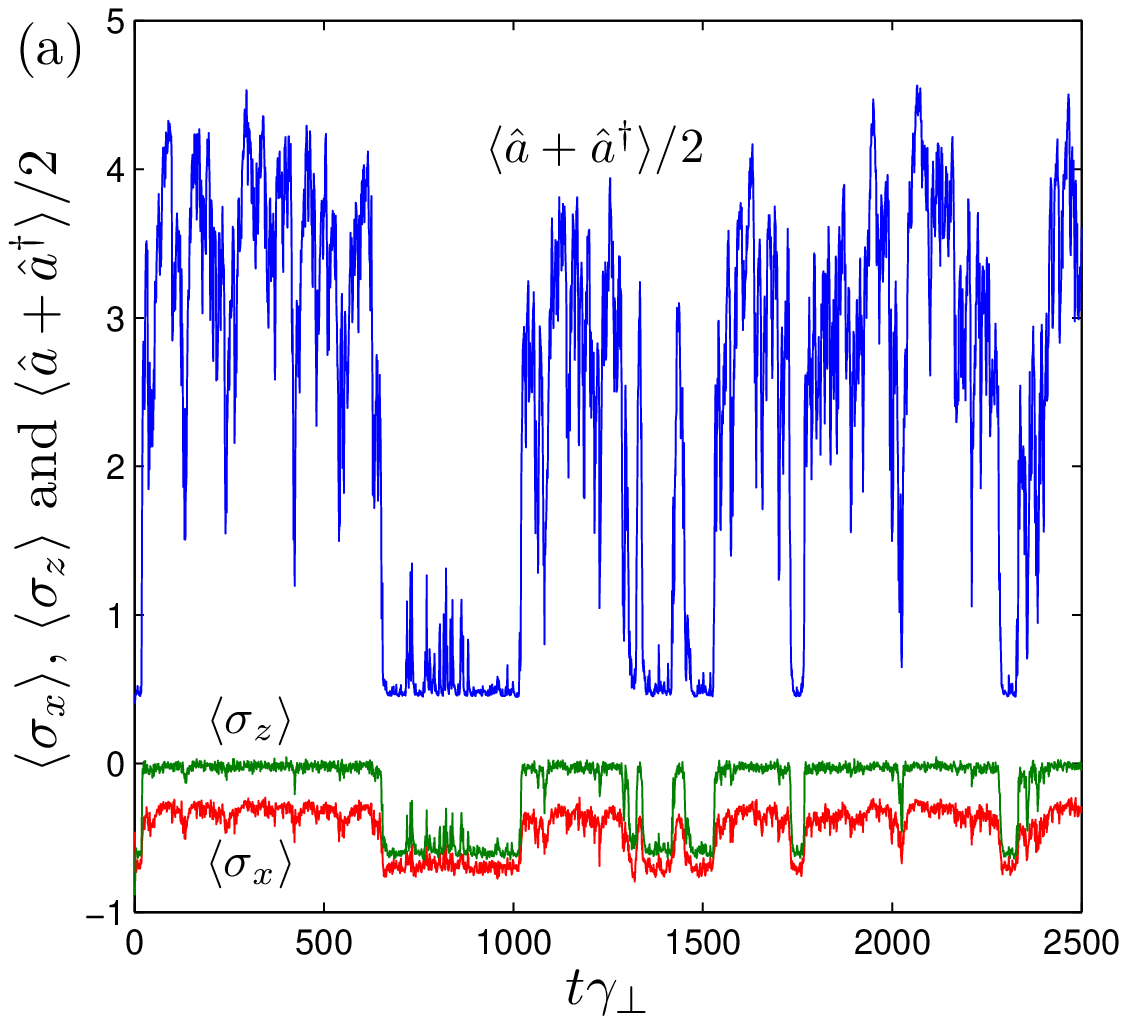}
\includegraphics[clip,width=0.41\textwidth]{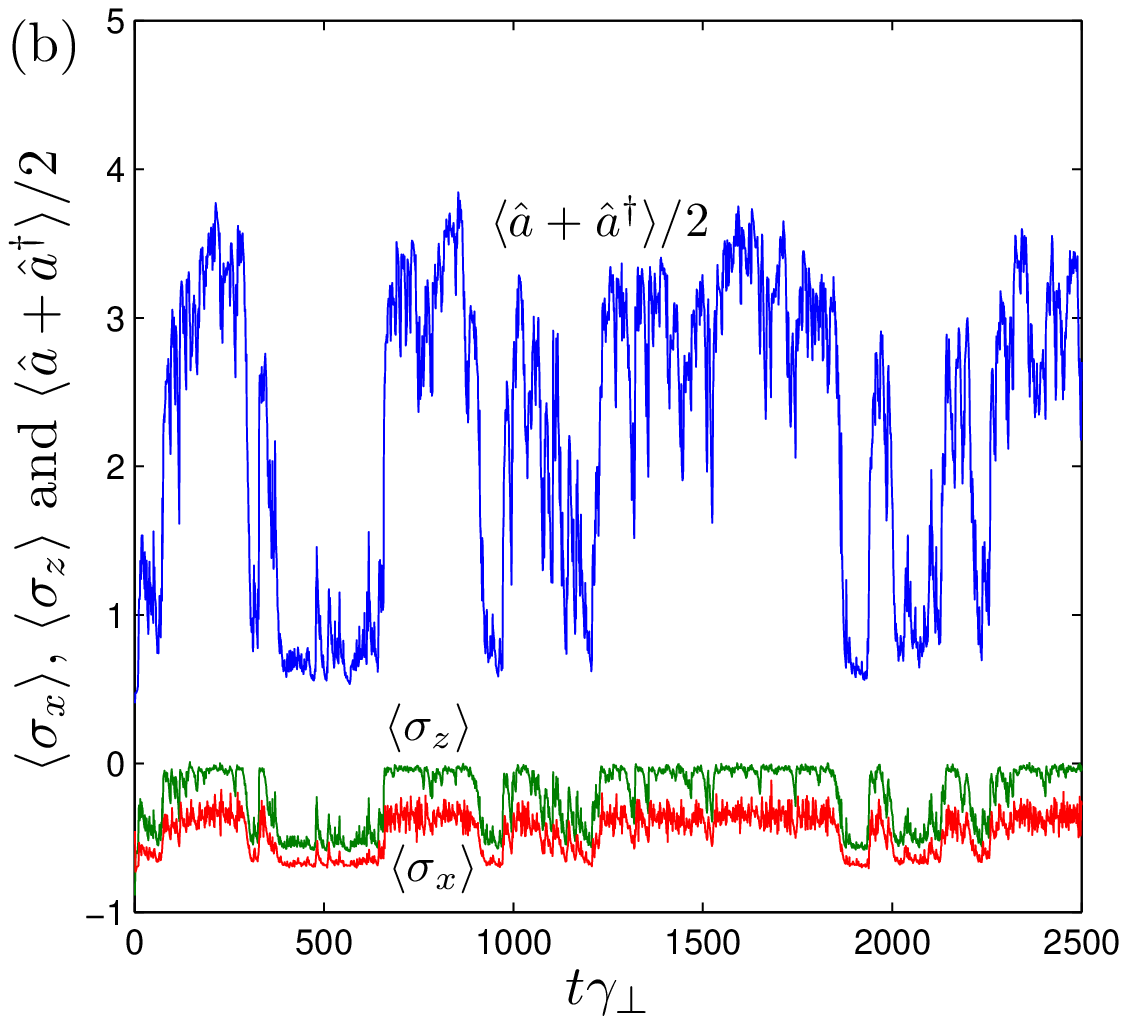}
\caption{Stochastic time evolution of $\langle\hat{a}+\hat{a}^\dag\rangle/2$ (blue), $\langle\sigma_z\rangle$ (green) and $\langle\sigma_x\rangle$ (red) for homodyne detection of the $x$-quadrature (a) and for homodyne detection of the $p$-quadrature (b).}\label{figure2}
\end{figure}

In figure~\ref{figure2} we show results for the time evolution of $\langle\hat{a}+\hat{a}^\dag\rangle/2$ and the expectation value of the Pauli operators $\sigma_x=\sigma+\sigma^\dag$ and $\sigma_z=[\sigma^\dag,\sigma]/2$ for a given realization of the measurement noise $\rmd W$. The system is seen to jump between two stable regions with different expectation values of the operators even though $\langle\hat{a}+\hat{a}^\dag\rangle/2$ fluctuates more in the upper region than it does in the lower region. The symmetry of (\ref{SME}) dictates that $\langle\sigma_y\rangle=\rmi\langle\sigma-\sigma^\dag\rangle$ and $-\rmi\langle\hat{a}-\hat{a}^\dag\rangle/2$ are both zero for the case of homodyne detection of the $x$-quadrature, while they fluctuate randomly around zero for the case of homodyne detection of the $p$-quadrature. Note that since the time average of $-\rmi\langle\hat{a}-\hat{a}^\dag\rangle/2$ is zero for both of the stable regions, our ability to distinguish the regions through a measurement of the $p$-quadrature relies on the fact that the time evolution of the state is different in the two regions.

\section{Model reduction}\label{3}

We now turn to the problem of deriving a simplified low-dimensional differential equation, which, ideally, contains the same dynamics as the full stochastic master equation (\ref{SME}). To do so, we first need to identify a suitable manifold onto which we can project the dynamics. Several manifold learning strategies have already been investigated in the literature \cite{saul,lin}, and here we use the method of local tangent space alignment (LTSA) \cite{ltsa}. An advantage of this method is that it optimizes the choice of low-dimensional space in local areas, which means that it is well suited to describe systems with more than one stable region. LTSA defines the manifold in terms of single points, but in order to perform the projection, we need a differentiable function, which relates the coordinates of all points in the low-dimensional space to the coordinates of the same points in the full space. This problem is solved by fitting a function to the points obtained from LTSA.

\subsection{Identification of the low-dimensional manifold}

In brief, the input to the LTSA algorithm is a set of $N$ vectors $x^{(i)}$, $i=1,2,\ldots,N$, sampled with noise from an unknown $d$-dimensional (nonlinear) manifold embedded in an $m$-dimensional space, where $m>d$, and the objective is to identify the underlying $d$-dimensional manifold. For a linear manifold this is done by computing the $d$-dimensional affine subspace, which minimizes the sum of the square of the errors between the original vectors and the vectors projected onto the affine subspace. To tackle the more general case, the LTSA procedure aligns local linear structures into a global nonlinear manifold, where the local structures are the affine subspaces obtained by applying the above procedure to subsets of the $N$ points. The $i$th subset is chosen as the $k$ nearest neighbours of the $i$th point (including the point itself), where $k$ is a number satisfying $d\ll k\ll N$. The output of the algorithm is the coordinates $\tau^{(i)}$ of the points in the $d$-dimensional space and an approximate map from the $d$-dimensional space to the full $m$-dimensional space, from which it is possible to compute corrected coordinates $\tilde{x}^{(i)}$ of the points in the $m$-dimensional space. The map is, however, only valid in small regions around each $\tau^{(i)}$, and it is not differentiable at all points.

In our case, we start from a set of density matrices sampled from the time evolution of the state of the system. Concretely, we choose the density matrix at times $t=501\textrm{ }\gamma_\bot^{-1}$, $t=502\textrm{ }\gamma_\bot^{-1}$, $\ldots$, $t=2500\textrm{ }\gamma_\bot^{-1}$ for the trajectory used to compute the results shown in figure~\ref{figure2}, and we choose $k=60$ as in \cite{hopkins}. These density matrices are transformed into real column vectors $x^{(i)}$, each with $14400$ elements, by concatenating the real part of the columns of the upper right triangular part including the diagonal followed by concatenation of the imaginary part of the columns of the upper right triangular part excluding the diagonal, i.e.,
\begin{eqnarray}\label{xi}
x^{(i)}_{n(n+1)/2}=\rho^{(i)}_{nn},\label{diagdel}\\
x^{(i)}_{m(m-1)/2+n}=\textrm{Re}(\rho^{(i)}_{nm}),\label{realdel}\\
x^{(i)}_{N(N+1)/2+(m-1)(m-2)/2+n}=\textrm{Im}(\rho^{(i)}_{nm}),\label{imagdel}
\end{eqnarray}
where $n=1,2,\ldots,N$ in (\ref{diagdel}) and $n=1,2,\ldots,m-1$ and $m=2,3,\ldots,N$ in (\ref{realdel}) and (\ref{imagdel}). This construction method ensures that every vector $x$ in the $m$-dimensional space corresponds to a Hermitian matrix. Furthermore, the LTSA algorithm ensures that the new points in the $m$-dimensional space are correctly normalized. There is, however, no guarantee that the constructed points correspond to positive semi-definite matrices, and here we rely on the ability of the time evolution equation to keep the state of the system within the physically acceptable region. Depending on the purpose of the reduced model, it is not necessarily optimal to minimize the projection error with respect to the dot product $(x^{(i)})^Tx^{(j)}$, and one could, for instance, consider to multiply the density matrix elements with different weight factors \cite{hopkins}. We note in particular that $(x^{(i)})^Tx^{(j)}$ is equal to $\Tr(\rho^{(i)}\rho^{(j)})$ if a factor of $\sqrt{2}$ is included on the right hand side of (\ref{realdel}) and (\ref{imagdel}), but we have avoided to do so in the following, because we obtain better results without the factor $\sqrt{2}$.

Having obtained a set of points $\tau^{(i)}$ in the low-dimensional space and the corresponding coordinates $\tilde{x}^{(i)}$ in the full space, we next construct a map from the low-dimensional space to the full space via fitting. We need to compute one fit for each of the $m=14400$ coordinates in the full space, and to make this procedure practical, we use the same fitting model for all the coordinates and choose this model to be linear in the fitting parameters, i.e., we assume a map of form $x=cf(\tau)$, where $x$ is a vector in the $m$-dimensional space, $c$ is an $m\times r$ matrix of fitting parameters, and $f$ is an $r\times1$ vector, whose elements $f_j$ are arbitrary functions of the coordinates $\tau=(\tau_1,\tau_2,\ldots,\tau_d)^T$ in the $d$-dimensional space. To ensure that $x$ is correctly normalized for all $\tau$, we choose $f_1(\tau)=1$ and minimize $\sum_i(\tilde{x}^{(i)}-cf(\tau^{(i)}))^T(\tilde{x}^{(i)}-cf(\tau^{(i)}))$ under the constraint $v^Tc=(1,0,\ldots,0)$, where $v$ is an $m\times1$ vector, whose $i$th entry is one if $i=n(n+1)/2$ for some $n\in\{1,2,\ldots,N\}$ and zero otherwise (i.e., $v^Tx=\Tr(\rho)$, where $\rho$ is the density matrix corresponding to the vector $x$). The result is
\begin{equation}\label{c}
c=\tilde{c}+\frac{1}{\sqrt{m}}v[(1,0,\ldots,0)-v^T\tilde{c}],
\end{equation}
where $\tilde{c}=[(z^Tz)^{-1}z^Ty]^T$, with $z_{ij}=f_j(\tau^{(i)})$ and $y_{ij}=\tilde{x}_j^{(i)}$, is the standard linear least squares result without constraints.

\subsection{Projection of the dynamics onto the identified manifold}

The result of the last subsection is a relation of form
\begin{equation}\label{rho}
\rho(\tau)=\sum_jc_jf_j(\tau_1,\tau_2,\ldots,\tau_d),
\end{equation}
where $c_j$ is the matrix obtained by applying the inverse of (\ref{diagdel}-\ref{imagdel}) to the $j$th column of $c$. To project the stochastic master equation onto the manifold defined by (\ref{rho}), we follow the derivation in \cite{hopkins}. We would like to interpret $\rmd\rho$ as a vector, but $\rmd\rho$ only transforms as a vector if $(\rmd W)^2=0$, i.e., if we use Stratonovich calculus, and we thus rewrite (\ref{SME}) into Stratonovich form $\rmd\rho=\underline{A}[\rho]\rmd t+B[\rho]\circ\rmd W$. Starting from a point $\rho(\tau)$ on the manifold, we then project $\rmd\rho(\tau)$ onto the tangent space of the manifold at that point, which is spanned by the $d$ vectors $\partial\rho(\tau)/\partial\tau_i$, using the dot product $\langle\rho_A,\rho_B\rangle\equiv\Tr(\rho_A\rho_B)$, i.e.,
\begin{eqnarray}
\fl\rmd\rho(\tau)=\sum_i\sum_j(g^{-1})_{ij}
\Tr\left\{\underline{A}[\rho(\tau)]\frac{\partial \rho(\tau)}{\partial \tau_j}\right\}\frac{\partial\rho(\tau)}{\partial\tau_i}\rmd t\nonumber\\
+\sum_i\sum_j(g^{-1})_{ij}\Tr\left\{B[\rho(\tau)]\frac{\partial \rho(\tau)}{\partial \tau_j}\right\}\frac{\partial\rho(\tau)}{\partial\tau_i}\circ\rmd W
\end{eqnarray}
where $g=g(\tau)$ is the metric tensor with elements
\begin{equation}
g_{ij}=\Tr\left(\frac{\partial\rho}{\partial \tau_i}\frac{\partial\rho}{\partial \tau_j}\right)=\sum_p\sum_q\frac{\partial f_p}{\partial\tau_i}\Tr(c_pc_q)\frac{\partial f_q}{\partial\tau_j}.
\end{equation}
Combining this relation with
\begin{equation}
\rmd\rho(\tau)=\sum_i\frac{\partial\rho(\tau)}{\partial \tau_i}\circ\rmd\tau_i,
\end{equation}
we obtain an expression for the time evolution of $\tau_i$
\begin{eqnarray}\label{dtaui}
\fl\rmd\tau_i=\sum_j(g^{-1})_{ij}\Tr\left\{\underline{A}[\rho(\tau)]
\frac{\partial\rho(\tau)}{\partial\tau_j}\right\}\rmd t\\
+\sum_j(g^{-1})_{ij}\Tr\left\{B[\rho(\tau)]\frac{\partial\rho(\tau)}{\partial \tau_j}\right\}\circ\rmd W.
\end{eqnarray}
To simplify the notation, we define vectors $v_1$ and $v_2$ with elements
\begin{eqnarray}
(v_1)_j&=\sqrt{2\kappa}\Tr[(\hat{a}^\dag\rme^{i\phi}+\hat{a}\rme^{-i\phi})c_j],\\
(v_2)_j&=2\kappa\textrm{Re}\{\Tr[(\hat{a}^\dag\rme^{i\phi}+\hat{a}\rme^{-i\phi})\hat{a}\rme^{-i\phi}c_j]\},
\end{eqnarray}
and matrices $M_g$, $M_H$ and $M_1$ with elements
\begin{eqnarray}
(M_g)_{ij}&=\Tr(c_ic_j),\\
(M_H)_{ij}&=2\textrm{Re}[\Tr(Mc_jc_i)]+2\gamma\textrm{Tr}(\sigma_-c_j\sigma_+c_i),\\
(M_1)_{ij}&=\sqrt{2\kappa}\Tr[(c_j\hat{a}^\dag\rme^{i\phi}-\hat{a}\rme^{-i\phi}c_j)c_i],
\end{eqnarray}
where $M=-\rmi H/\hbar-\kappa\hat{a}^\dag\hat{a}-\kappa\hat{a}^2\rme^{-2i\phi}
-\gamma\sigma_+\sigma_-$. Finally, we insert $\underline{A}[\rho(\tau)]$ and $B[\rho(\tau)]$ into (\ref{dtaui}) and convert back to it\^o form to obtain
\begin{equation}\label{dtaui2}
d\tau_i=a_i[\tau]dt+b_i[\tau]dW,
\end{equation}
where
\begin{eqnarray}
\fl a_i[\tau]=\frac{1}{2}v_1^Tfb_i[\tau]+\sum_j(g^{-1})_{ij}
\left(\frac{\partial f^T}{\partial \tau_j}M_Hf+v_2^Tf\frac{\partial f^T}{\partial \tau_j}M_gf\right)
+\frac{1}{2}\sum_j\sum_k(g^{-1})_{ij}\nonumber\\
\times\left(\frac{\partial^2f^T}{\partial \tau_k\partial \tau_j}M_1f
+\frac{\partial f^T}{\partial \tau_j}M_1\frac{\partial f}{\partial \tau_k}
-v_1^T\frac{\partial f}{\partial \tau_k}\frac{\partial f^T}{\partial \tau_j}M_gf-v_1^Tf\frac{\partial^2f^T}{\partial \tau_k\partial \tau_j}M_gf\right)b_k[\tau]\nonumber\\
-\frac{1}{2}\sum_j\sum_k\sum_q
(g^{-1})_{ij}\left(\frac{\partial^2f^T}{\partial \tau_k\partial \tau_j}M_g\frac{\partial f}{\partial \tau_q}+\frac{\partial f^T}{\partial \tau_j}M_g\frac{\partial^2f}{\partial \tau_k\partial \tau_q}\right)b_k[\tau]b_q[\tau]
\end{eqnarray}
and
\begin{equation}
b_i[\tau]=\sum_j(g^{-1})_{ij}\left(\frac{\partial f^T}{\partial \tau_j}M_1f-v_1^Tf\frac{\partial f^T}{\partial \tau_j}M_gf\right).
\end{equation}
Integrating the low-dimensional equation (\ref{dtaui2}), we can now approximately predict the time evolution of $\rho$ through (\ref{rho}).

\section{Performance of the reduced models}\label{4}

\begin{figure}
\flushright
\includegraphics[clip,width=0.4\textwidth]{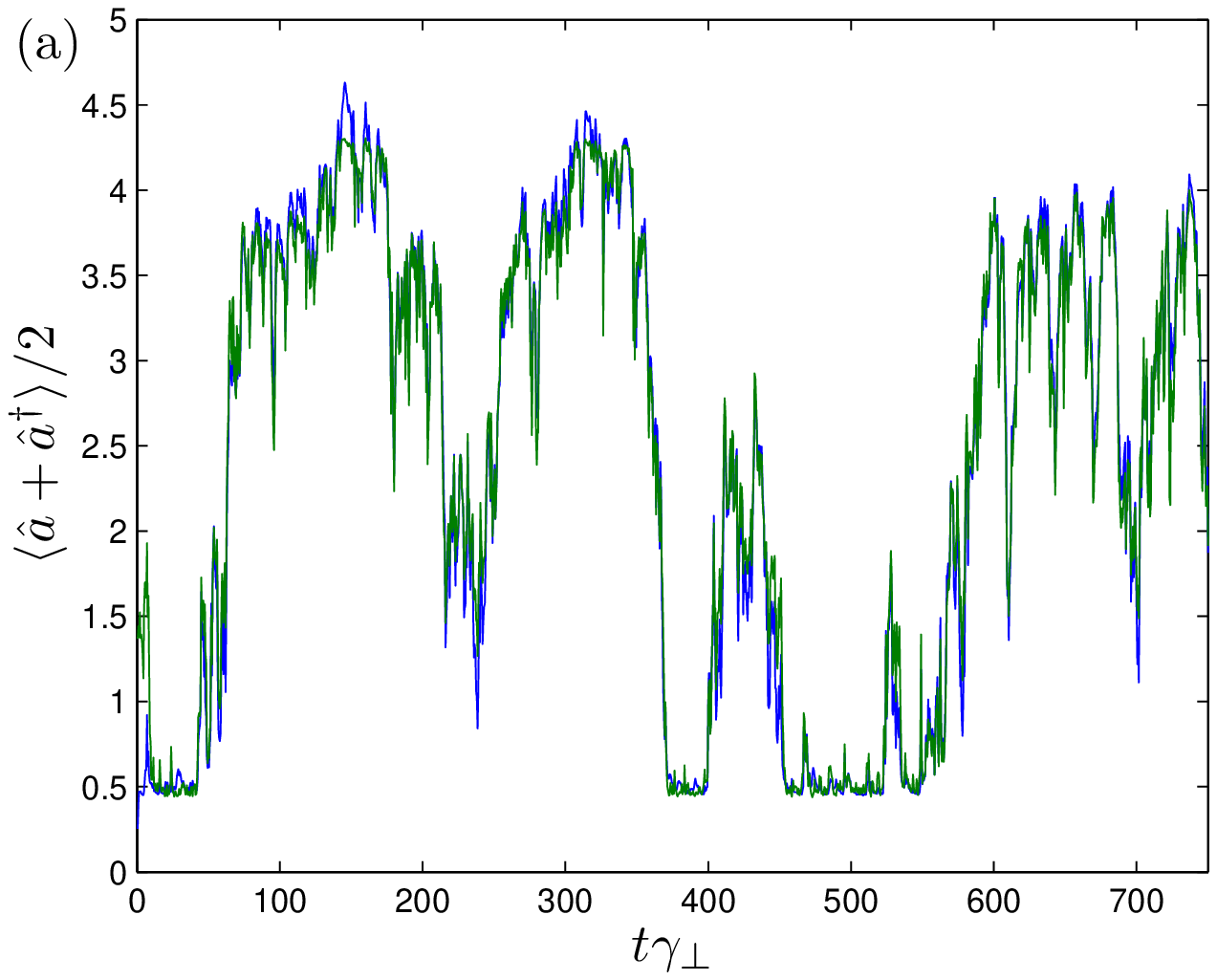}\hspace{0.02\textwidth}
\includegraphics[clip,width=0.4\textwidth]{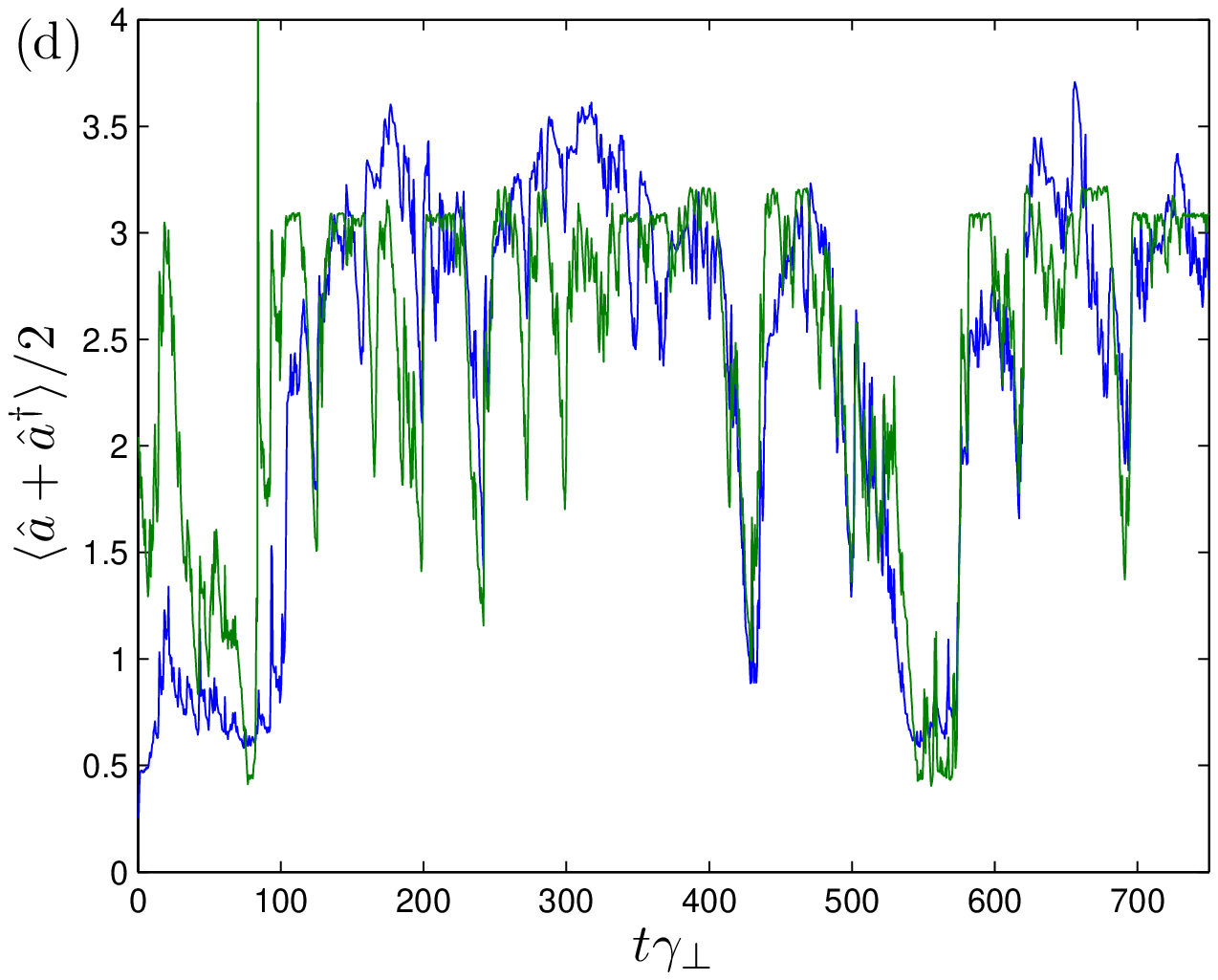}\\
\includegraphics[clip,width=0.4\textwidth]{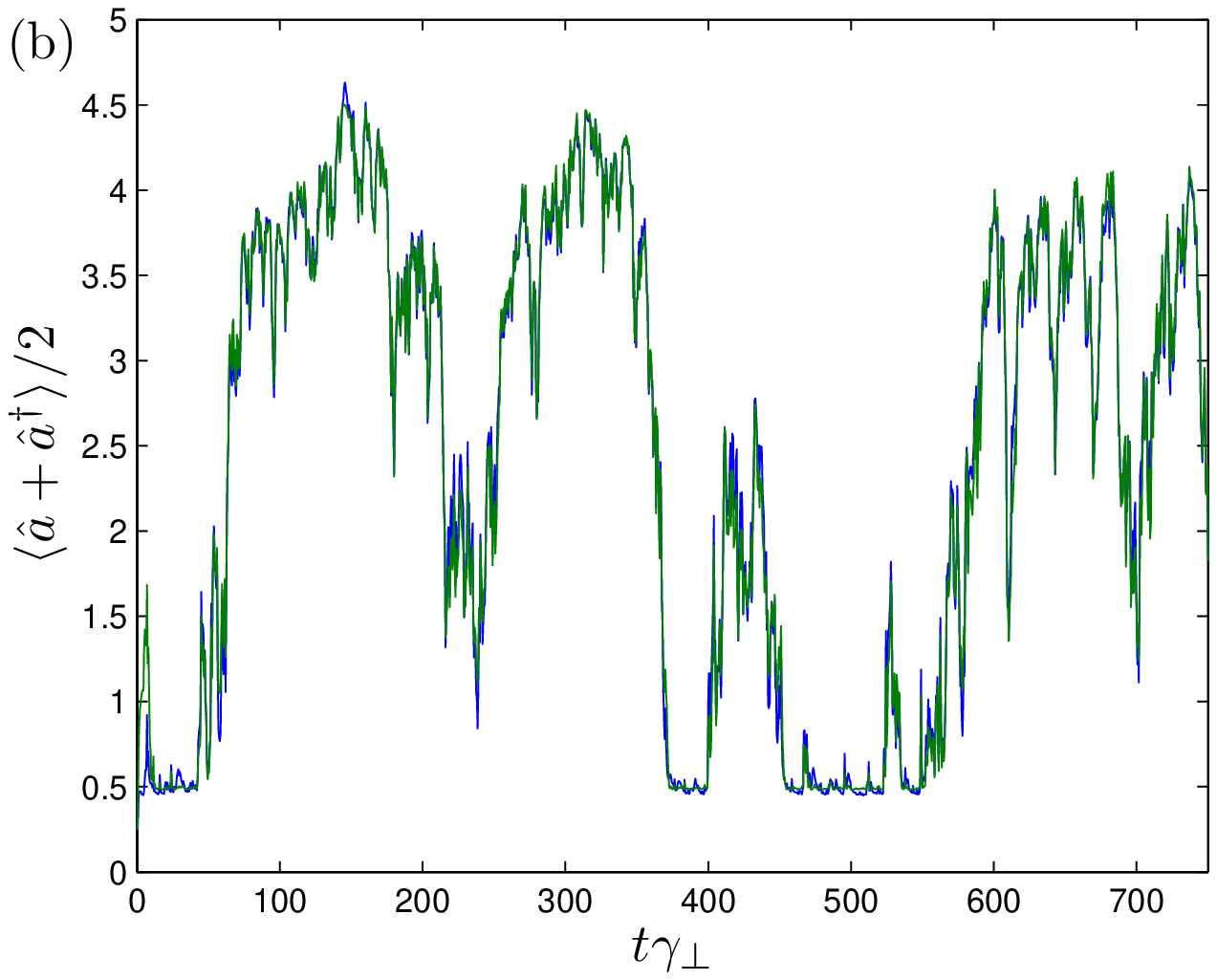}\hspace{0.02\textwidth}
\includegraphics[clip,width=0.4\textwidth]{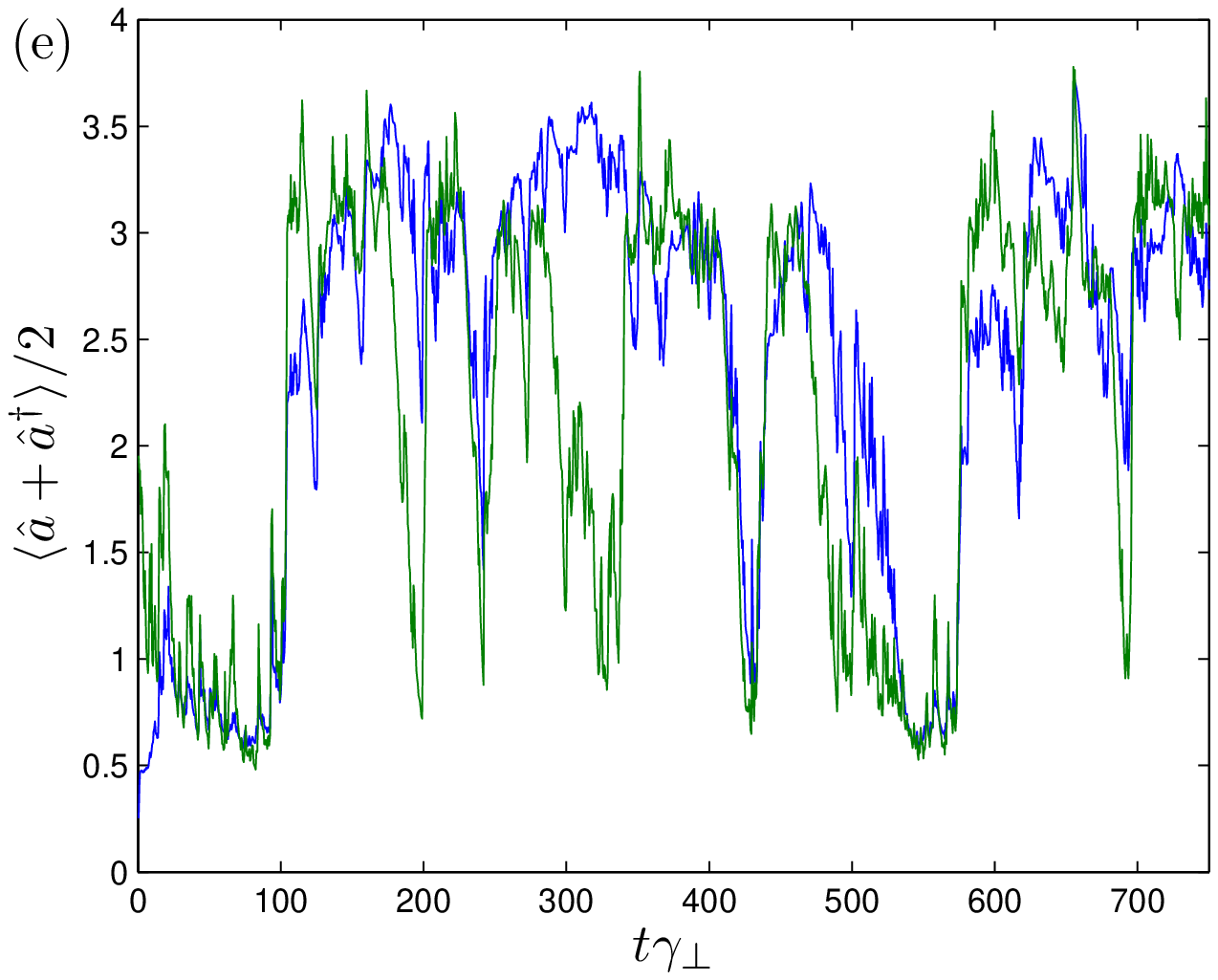}\\
\includegraphics[clip,width=0.4\textwidth]{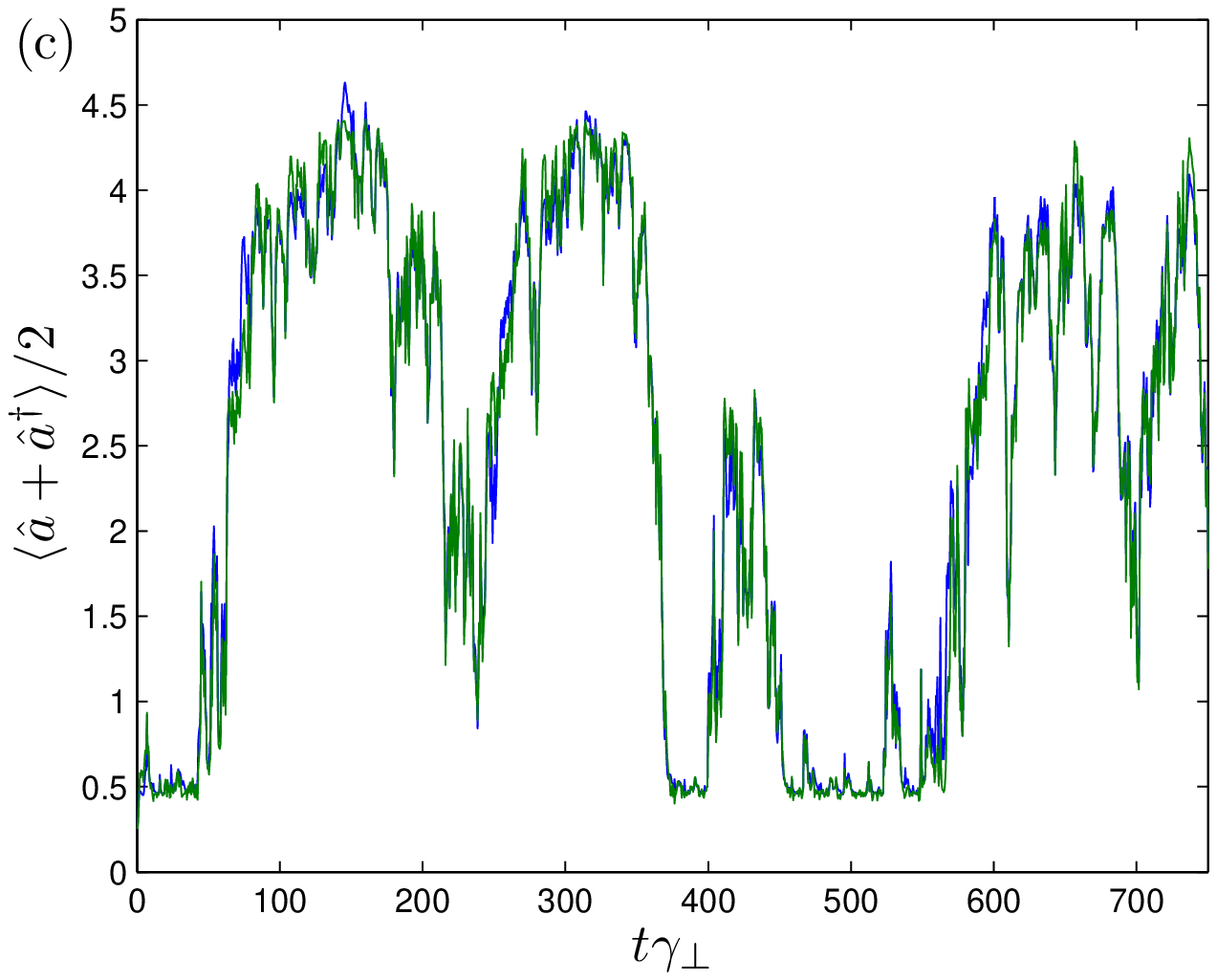}\hspace{0.02\textwidth}
\includegraphics[clip,width=0.4\textwidth]{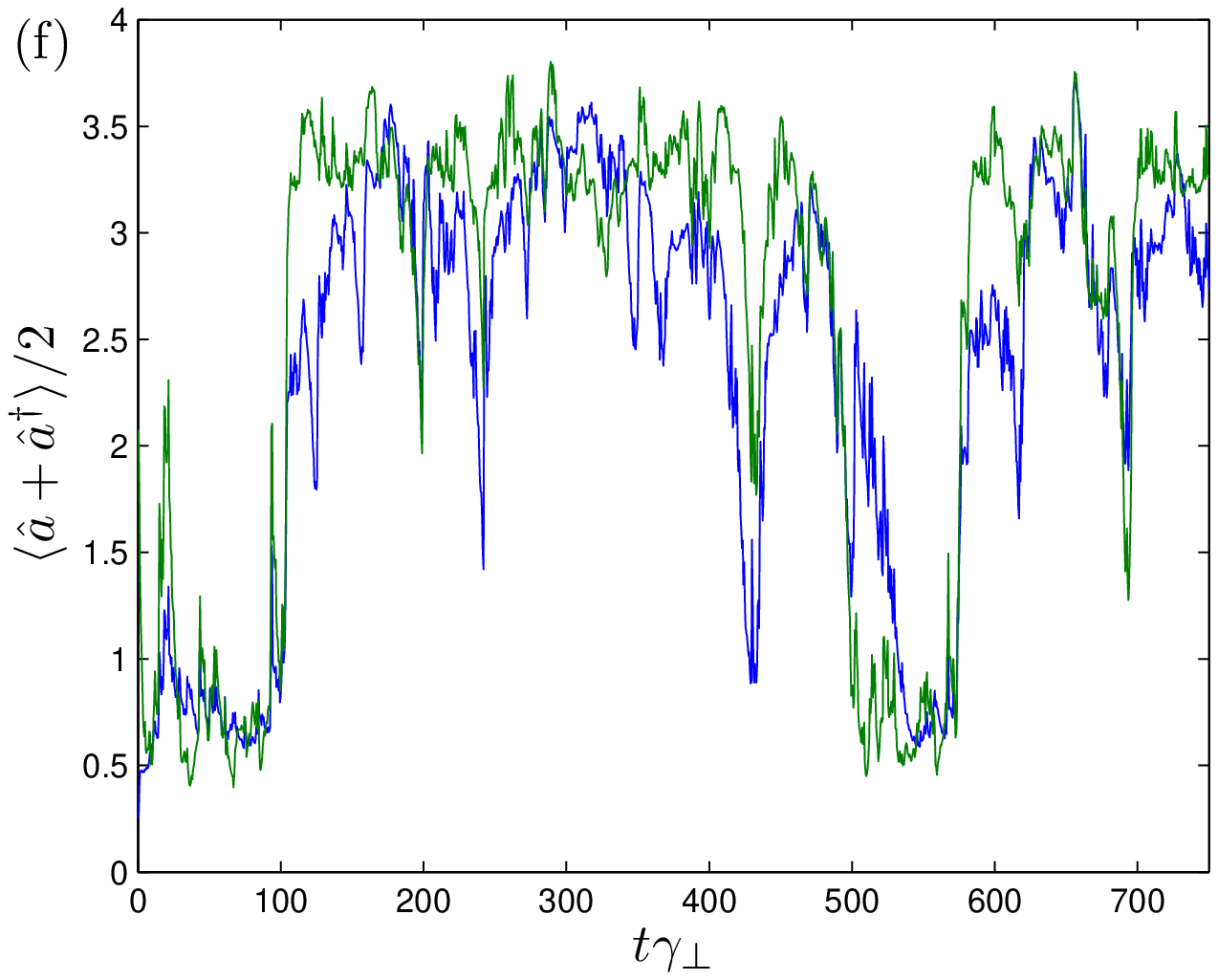}
\caption{Time evolution of $\langle\hat{a}+\hat{a}^\dag\rangle/2$ obtained from the full stochastic master equation (blue) and from various reduced models (green) assuming the same photo current. For (a) and (d) the fitting model is a second order polynomial in two dimensions, i.e., $f=(1,\tau_1,\tau_2,\tau_1^2,\tau_1\tau_2,\tau_2^2)^T$, for (b) and (e) it is a fourth order polynomial in two dimensions, and for (c) and (f) it is a second order polynomial in four dimensions. The left figures are for homodyne detection of the $x$-quadrature, and the right figures are for homodyne detection of the $p$-quadrature.}\label{figure3}
\end{figure}

Since we would like to use the reduced models to predict the state of the system in a feedback scheme, we should check the performance of the reduced models by generating a realistic photo current using the full stochastic master equation and then use that photo current to integrate the reduced models. Examples of this procedure, using polynomials as fitting models, are provided in figure~\ref{figure3}. We have plotted $\langle\hat{a}+\hat{a}^\dag\rangle/2$, because this is the quantity we need to predict in the feedback scheme proposed in the next section. For the case of homodyne detection of the $p$-quadrature we have used a weighted linear least squares method to compute $\tilde{c}$ in (\ref{c}) to reduce the effect of outliers. The precise initial state of the reduced models is not important, because the observed value of the photo current quickly drags the reduced model to the correct state, and we have thus chosen $\tau=(0,0,\ldots,0)^T$ (the average of $\tau^{(i)}$) for simplicity. In case of instability, we have reset the reduced model to $\tau=(0,0,\ldots,0)^T$ whenever the program returns a non-determined value of $\tau$.

The figure shows that simple low-dimensional models are able to provide accurate predictions for the value of $\langle\hat{a}+\hat{a}^\dag\rangle/2$ for the case of homodyne detection of the $x$-quadrature. For homodyne detection of the $p$-quadrature, the two-dimensional model obtained by using a second order polynomial as fitting model is observed to largely reproduce the time evolution of $\langle\hat{a}+\hat{a}^\dag\rangle/2$, but the details differ. In particular, the reduced model does not reproduce the highest values of $\langle\hat{a}+\hat{a}^\dag\rangle/2$. To improve the agreement between the full and the reduced model, one could try to increase the order of the fitting polynomial or use a higher-dimensional model. For a fourth order polynomial in two dimensions, the reduced model is able to reach the highest values of $\langle\hat{a}+\hat{a}^\dag\rangle/2$ and the value of $\langle\hat{a}+\hat{a}^\dag\rangle/2$ in the lower state is also more accurate, but the model predicts false jumps to the lower state, which is undesirable in a feedback scheme. False jumps are not observed in the figure for the four-dimensional model with a second order polynomial as fitting model, and we thus use this latter model to predict the state of the system in the next section.

The reason why we obtain very accurate results for homodyne detection of the $x$-quadrature is that we use the actual photo current $\rmd y$ obtained from the full stochastic master equation to drive the reduced model. This means that the back action of the measurement on the system is large whenever the predicted value of the measured quadrature differs significantly from the value obtained from the full stochastic master equation. It is, in fact, a much harder test of the performance of the reduced models to check whether they are able to predict the value of quantities that are not related in a simple way to the observed quadrature such as the value of $\langle\hat{a}+\hat{a}^\dag\rangle/2$ for homodyne detection of the $p$-quadrature. The above results thus indicate that the low-dimensional models actually capture most of the full dynamics of the system. As expected, we also find that the reduced models for homodyne detection of the $p$-quadrature provide more accurate results for $-i\langle\hat{a}-\hat{a}^\dag\rangle/2$ than for $\langle\hat{a}+\hat{a}^\dag\rangle/2$.

\section{Stabilization of one attractor through feedback control}\label{5}

A natural feedback scheme to hold the system within one of the stable regions is to increase or decrease the intensity of the drive laser depending on whether the value of $\langle\hat{a}+\hat{a}^\dag\rangle/2$ is below or above some suitably chosen target value $x_0$. This is achieved by adding a proportional feedback term
\begin{equation}\label{sfb}
\rmd\rho_{\textrm{fb}}=s_pe(t)[\hat{a}-\hat{a}^\dag,\rho]\gamma_\bot\rmd t,\qquad
e(t)\equiv\langle\hat{a}+\hat{a}^\dag\rangle/2-x_0,
\end{equation}
to the stochastic master equation (\ref{SME}), where $s_p$ is a parameter determining the strength of the feedback. Integration of the resulting equation confirms that the feedback term has the desired effect, and it is possible to decrease the standard deviation of $e(t)$ to a value, which is very small compared to typical fluctuations of $\langle\hat{a}+\hat{a}^\dag\rangle/2$ without feedback. The question is now whether this is still the case if we use a reduced model (with the feedback term included) to predict the value of $e(t)$.

\begin{figure}
\flushright
\includegraphics[clip,width=0.4\textwidth]{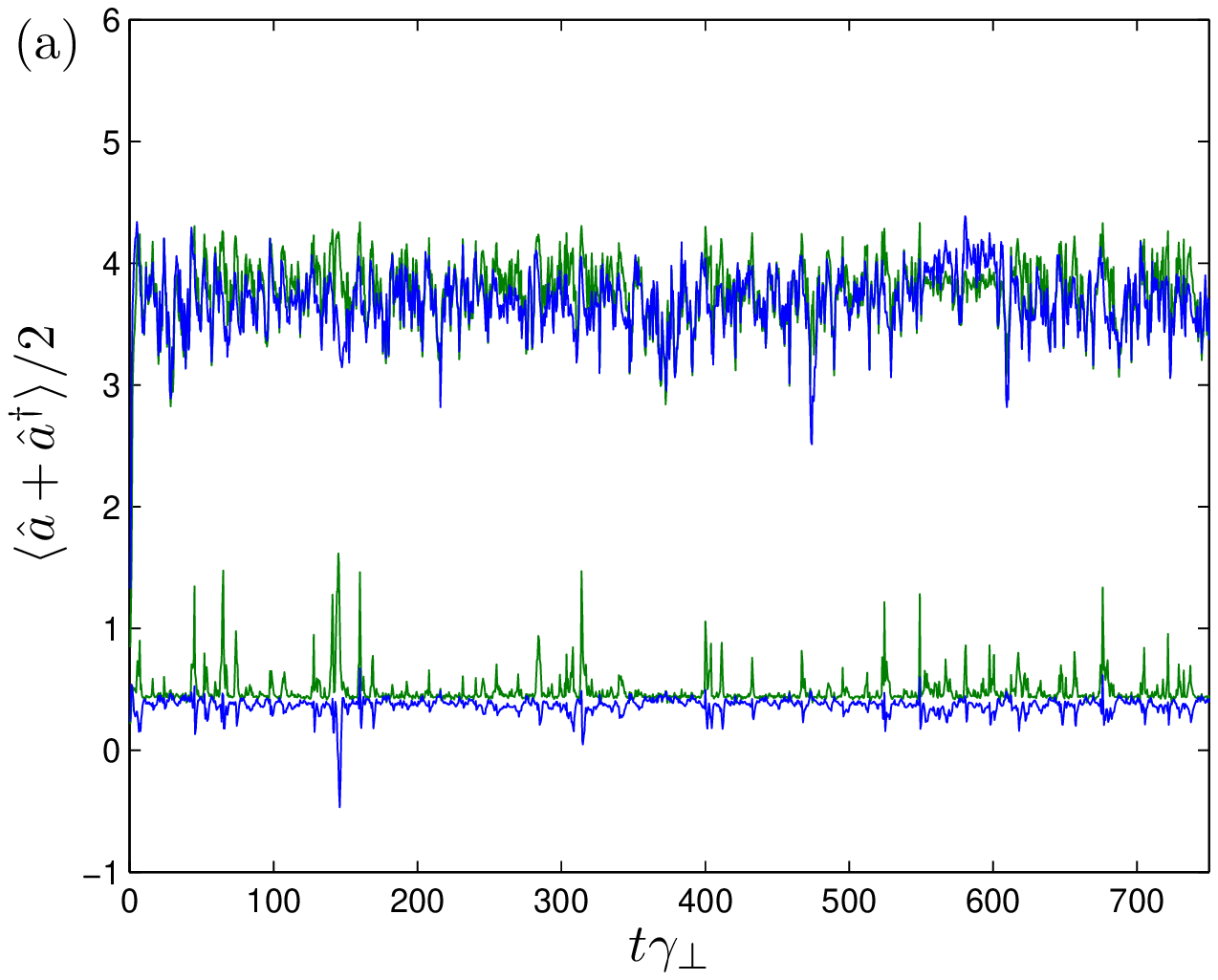}\hspace{0.02\textwidth}
\includegraphics[clip,width=0.4\textwidth]{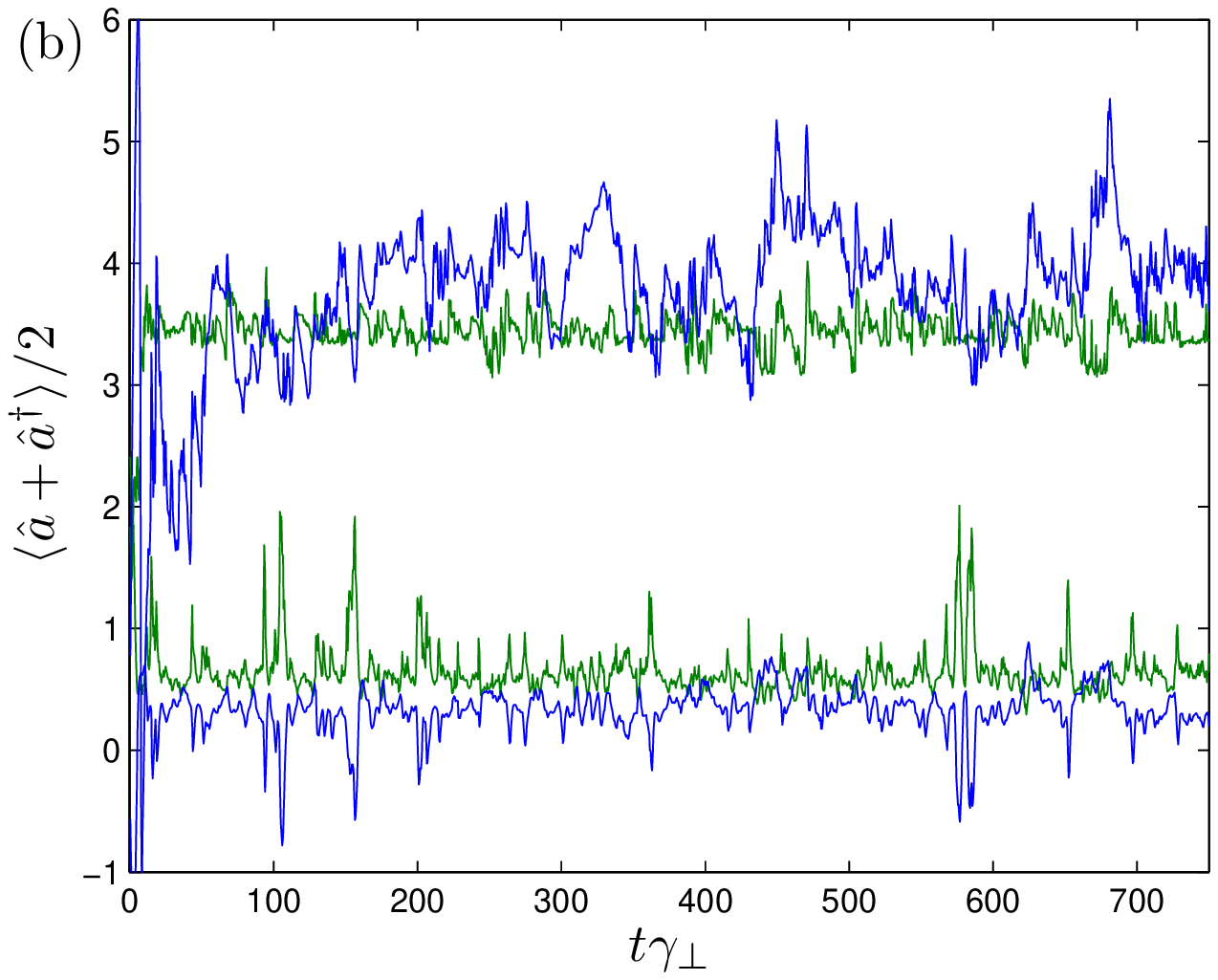}
\caption{Time evolution of $\langle\hat{a}+\hat{a}^\dag\rangle/2$ obtained from the full stochastic master equation (blue) for homodyne detection of the $x$-quadrature (a) and homodyne detection of the $p$-quadrature (b) for the same noise realization as in figure~\ref{figure3}, but with the feedback term in (\ref{sfb}) included. The error $e(t)$ is computed from the value of $\langle\hat{a}+\hat{a}^\dag\rangle/2$ predicted by the four-dimensional reduced model with a second order polynomial as fitting model (green), and as in the last section we have used the photo current obtained from the full stochastic master equation to integrate the low-dimensional model. In (a), $x_0=3.8$ and $s_p=0.75$ for the upper curves and $x_0=0.4$ and $s_p=1$ for the lower curves. In (b), $x_0=3.5$ and $s_p=0.75$ for the upper curves and $x_0=0.5$ and $s_p=1$ for the lower curves.}\label{figure4}
\end{figure}

Example trajectories are shown in figure~\ref{figure4}. Comparing these trajectories to those in figure~\ref{figure3}, it is apparent that the feedback term affects the time evolution, and that the system stays close to the upper or the lower stable region for the considered values of $x_0$. For the upper stable region, $\langle\hat{a}+\hat{a}^\dag\rangle/2$ is seen to oscillate with a relatively large amplitude, which reflects the fact that the upper stable region is relatively broad as observed in figure~\ref{figure2}. For the lower stable region, $\langle\hat{a}+\hat{a}^\dag\rangle/2$ is roughly constant except for sudden spikes. By choosing a value of $x_0$, which is slightly below the average value of $\langle\hat{a}+\hat{a}^\dag\rangle/2$ predicted by the reduced model, we can ensure that the feedback term almost always acts to decrease $\langle\hat{a}+\hat{a}^\dag\rangle/2$. This keeps the full model away from the transition region, and as seen in the figure the spikes tend to point towards negative values of $\langle\hat{a}+\hat{a}^\dag\rangle/2$ rather than towards larger positive values of $\langle\hat{a}+\hat{a}^\dag\rangle/2$ as observed for higher values of $x_0$.

For the case of homodyne detection of the $p$-quadrature, we note that the predictions of the reduced model for $x_0=3.5$ fluctuate less than the results obtained from the full stochastic master equation, while the fluctuations of the full and the reduced model are approximately the same for the case of homodyne detection of the $x$-quadrature. This is because we use the reduced model to evaluate the error $e(t)$, which means that the feedback term always acts to reduce $e(t)$ for the low-dimensional model. For the full model, on the other hand, $e(t)$ may have the opposite sign, in which case the full model is pushed away from $x_0$. The resulting change in the photo current drives the reduced model in the same direction as the full model, but this mechanism is more efficient in the case of homodyne detection of the $x$-quadrature than for homodyne detection of the $p$-quadrature.

\begin{figure}
\flushright
\includegraphics[clip,width=0.369\textwidth]{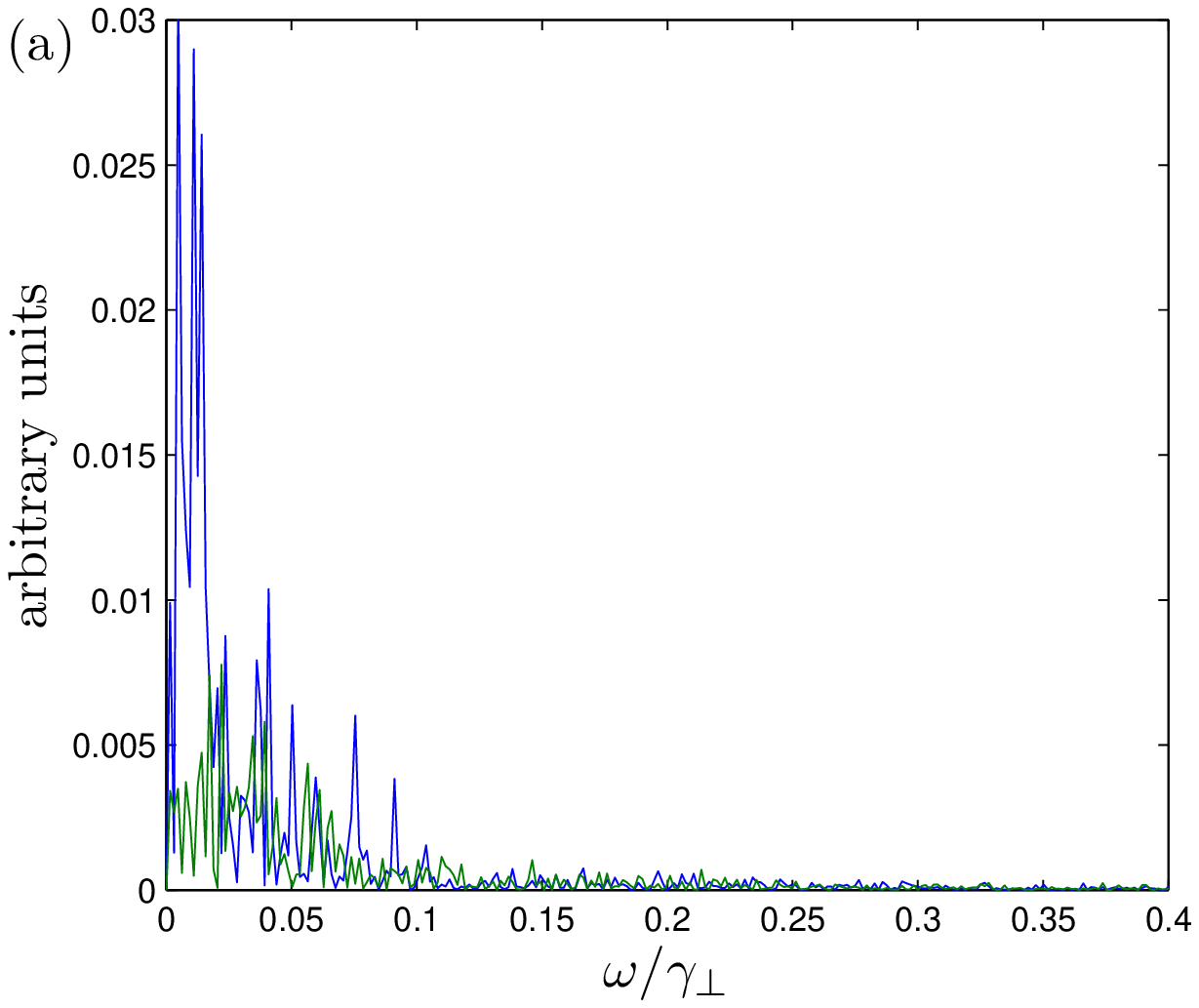}\hspace{0.02\textwidth}
\includegraphics[clip,width=0.431\textwidth]{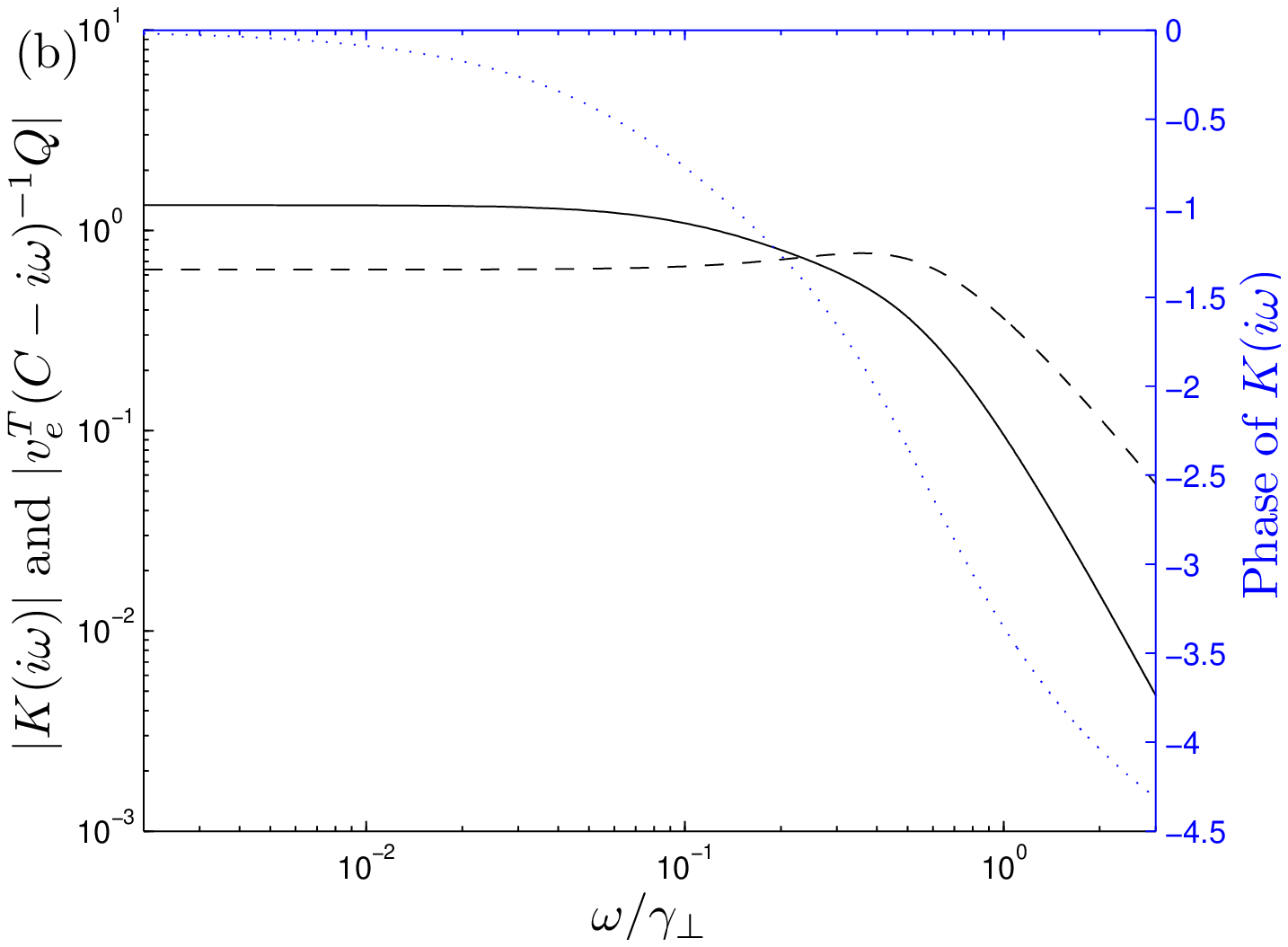}
\caption{(a) Power spectrum of a trajectory corresponding to the one in figure~\ref{figure3}(f), but integrated to $t\gamma_\bot=4000$. The blue curve is the power spectrum of $\langle\hat{a}+\hat{a}^\dag\rangle/2$ obtained from the full stochastic master equation, the green curve is the power spectrum of the difference between the reduced and the full model, and $\omega$ is the angular frequency. In both cases we have subtracted the mean value before computing the power spectrum. (b) Plots of $|v_e^T(C-\rmi\omega)Q|$ (dashed) and of the norm (solid) and the phase (dotted) of the loop transfer function $K(\rmi\omega)$.}\label{figure5}
\end{figure}

Discrepancies between the full model and the reduced model lead to a feedback of noise into the system, and even though a large value of $s_p$ may reduce the variance of $e(t)$ for the low-dimensional model, we observe that the predictions for $\langle\hat{a}+\hat{a}^\dag\rangle/2$ obtained from the full model fluctuate over a range that is broader than the distance between the upper and the lower stable region if $s_p$ is chosen too large. The power spectrum of the time evolution of $\langle\hat{a}+\hat{a}^\dag\rangle/2$ for a trajectory computed from the full stochastic master equation with homodyne detection of the $p$-quadrature and the power spectrum of the difference between the predictions of the reduced model and the full model in figure~\ref{figure5}(a) show that the relative error of the reduced model in predicting $\langle\hat{a}+\hat{a}^\dag\rangle/2$ is smaller at low frequencies. This appears because the reduced model is able to predict quantum jumps of the system but does not capture the details of the dynamics in the stable regions, and it suggests that it might be an advantage to mainly feed back the low frequency behaviour, which can be achieved by adding an integral term to the controller
\begin{equation}\label{fb}
\rmd\rho_{\textrm{fb}}=\left[s_pe(t)+s_i\int_0^t
\exp\left(-\zeta(t-t')\right)e(t')dt'\right]
[\hat{a}-\hat{a}^\dag,\rho]\gamma_\bot\rmd t,
\end{equation}
where $s_i$ and $\zeta$ are constants.

Rough estimates of reasonable choices of $s_p$, $s_i$ and $\zeta$ can be obtained as follows. If the measurement noise is turned off by setting $dW=0$ in the full stochastic master equation, the state of the system decays to steady state, which is an incoherent mixture of the states corresponding to the upper and the lower stable regions. (Note that $dW$ in equation (\ref{dtaui2}) may be nonzero, since we still require that the the photo current is the same for the reduced and the full model.) The behaviour of the system when a small input drive field term $\rmd\rho_{u}=u(t)[\hat{a}-\hat{a}^\dag,\rho]\gamma_\bot\rmd t$ is added can then be investigated by linearizing (\ref{SME}) and (\ref{dtaui2}) around the steady state point, which leads to an equation of form
\begin{equation}\label{linear}
\frac{\rmd}{\rmd t}\left[\begin{array}{c}\delta\tau\\ \delta\rho\end{array}\right]=C\left[\begin{array}{c}\delta\tau\\ \delta\rho\end{array}\right]+Qu(t),
\end{equation}
where $\delta\tau$ is the deviation of $\tau$ from the steady state value, $\delta\rho$ is a vector of the deviations of the density matrix elements from their steady state values with one diagonal element omitted, $C$ is a $14403\times14403$ matrix and $Q$ is a $14403\times1$ vector. Choosing $x_0$ to be the steady state value of $\langle\hat{a}+\hat{a}^\dag\rangle/2$ obtained from the reduced model, we can write the error as
\begin{equation}
e(t)=v_e^T\left[\begin{array}{c}\delta\tau\\ \delta\rho\end{array}\right],
\end{equation}
where $v_e$ is a $14403\times1$ vector for which all but the first four elements are zero. Integrating the linear equation (\ref{linear}), we have an expression for the time evolution of $e(t)$, which we insert into (\ref{fb}) to obtain the feedback term for a given input $u(t)$. In a closed loop setting, the feedback term is used as input, and the system behaviour is thus characterized by the loop transfer function
\begin{equation}\label{K}
K(s)=-\left(s_p+\frac{s_i}{\zeta+s}\right)v_e^T(C-s)^{-1}Q,
\end{equation}
which is the Laplace transform of the coefficient in square brackets in (\ref{fb}) divided by the Laplace transform of $u(t)$. For $s=i\omega$, the norm of $K$ is the loop gain at angular frequency $\omega$, and according to the power spectra in figure~\ref{figure5}(a) this should be close to zero for angular frequencies above approximately $2\pi\times0.02~\gamma_\bot=0.126~\gamma_\bot$. The norm of $v_e^T(C-i\omega)^{-1}Q$ is determined completely by the system and is plotted in figure~\ref{figure5}(b) for the case of homodyne detection of the $p$-quadrature. Since this factor is substantially different from zero for a range of angular frequencies above $2\pi\times0.02~\gamma_\bot$, we choose $s_p=0$. To ensure that $|s_i(\zeta+i\omega)^{-1}|$ has a large negative derivative for $\omega\approx2\pi\times0.02~\gamma_\bot$, we set the angular cross-over frequency to $\zeta=2\pi\times0.02~\gamma_\bot$. Finally, we choose $s_i=\sqrt{2}\zeta(|v_e^T(C-i\zeta)^{-1}Q|)^{-1}=0.26~\gamma_\bot$ such that $|K(i\zeta)|=1$. The resulting norm and phase of the loop transfer function are also plotted in figure~\ref{figure5}(b). The phase is seen to be well above $-\pi$ at the cross-over frequency, and we thus expect the feedback to be stable. One should, however, not read too much into (\ref{K}) as the above derivation is very crude. A similar analysis for the case of homodyne detection of the $x$-quadrature leads to the parameters $s_p=0$, $s_i=0.51~\gamma_\bot$ and $\zeta=2\pi\times0.1~\gamma_\bot$ and suggests that the feedback may be unstable for $s_i\gtrsim0.7~\gamma_\bot$.

\begin{figure}
\flushright
\includegraphics[clip,width=0.40\textwidth]{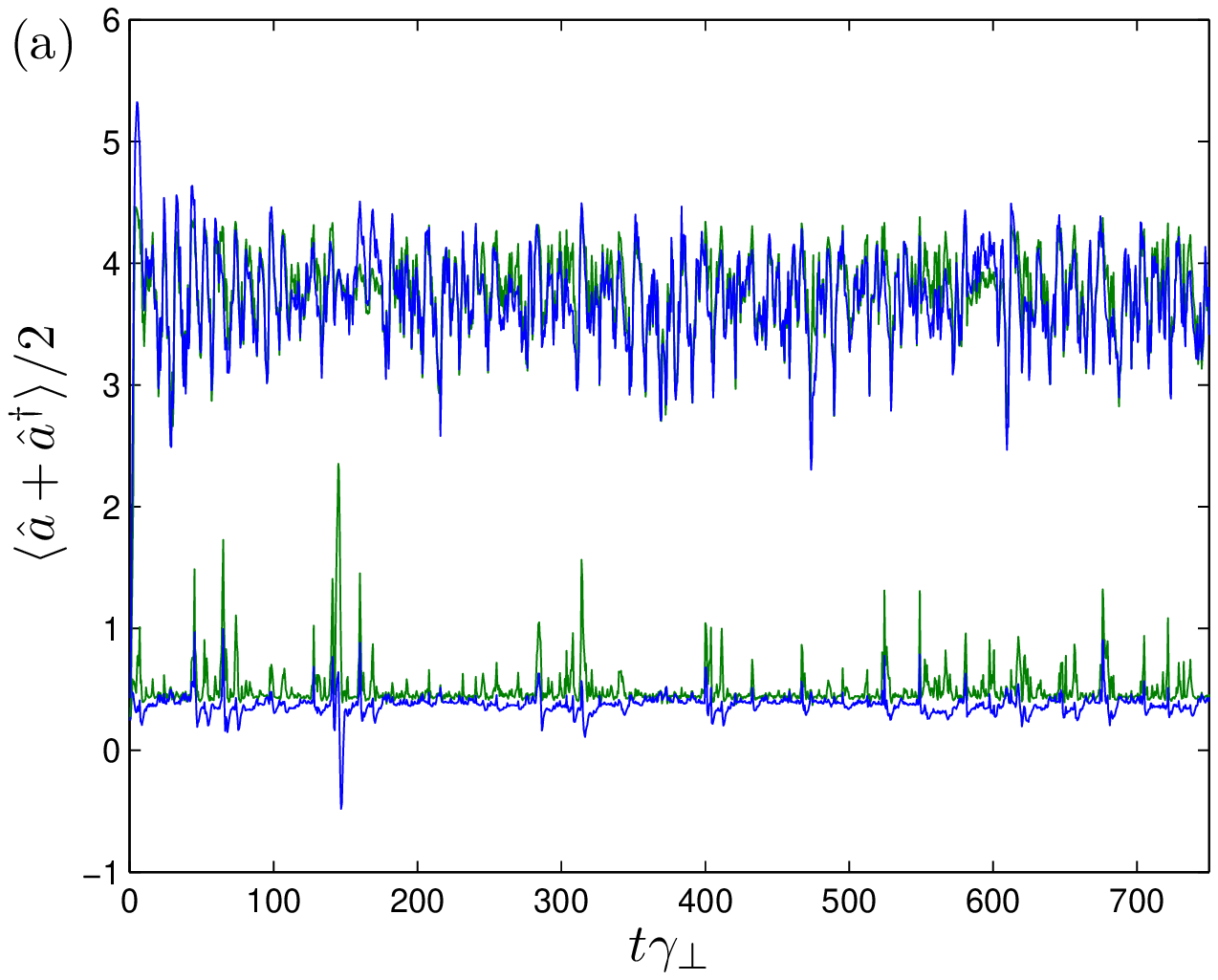}\hspace{0.02\textwidth}
\includegraphics[clip,width=0.40\textwidth]{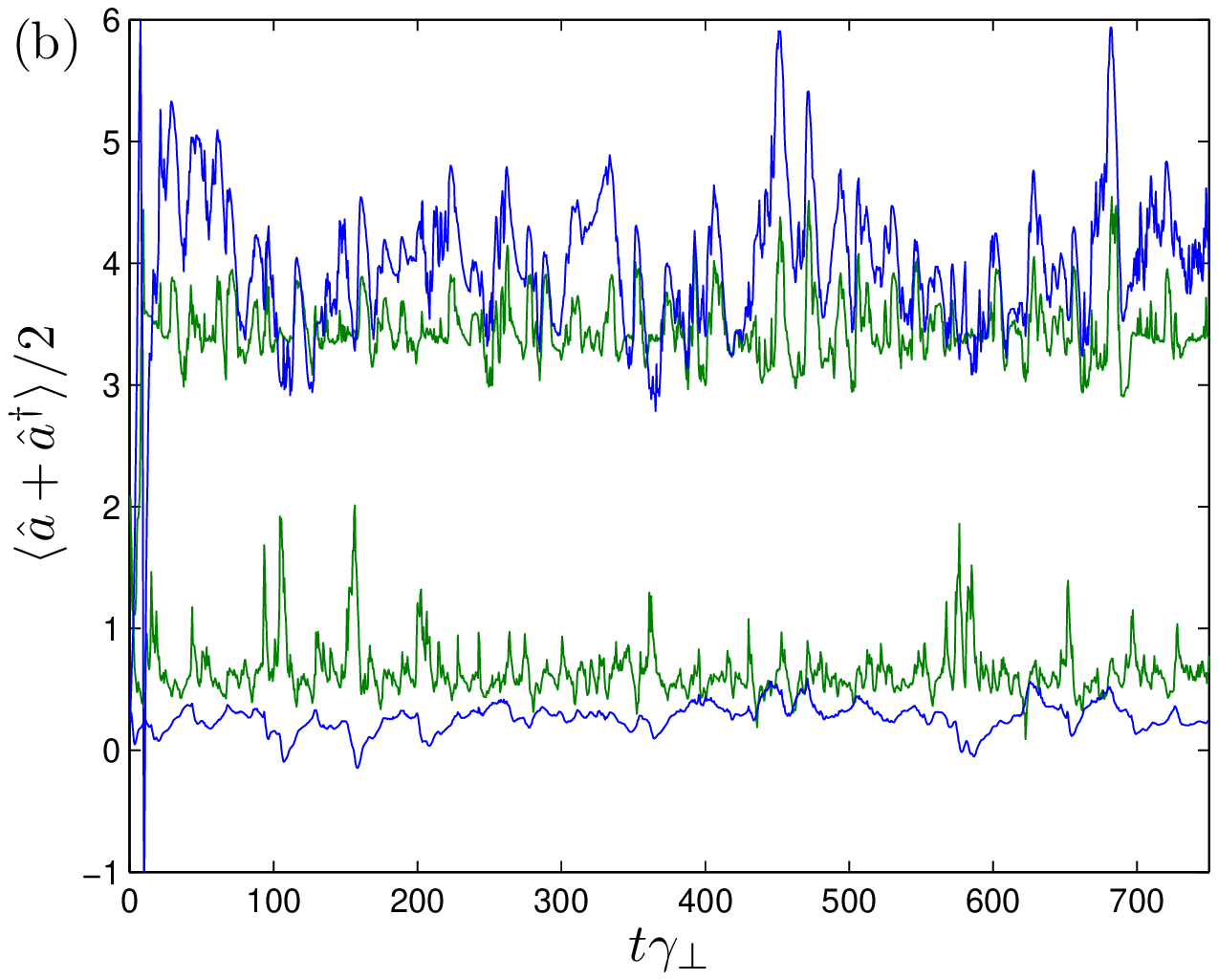}
\caption{Same as in figure~\ref{figure4}, but for the feedback term in (\ref{fb}). The parameters are $s_p=0$, $s_i=0.51~\gamma_\bot$ and $\zeta=2\pi\times0.1~\gamma_\bot$ for the case of homodyne detection of the $x$-quadrature and $s_p=0$, $s_i=0.15~\gamma_\bot$ and $\zeta=2\pi\times0.02~\gamma_\bot$ for the case of homodyne detection of the $p$-quadrature.}\label{figure6}
\end{figure}

Keeping $s_p$ and $\zeta$ fixed and searching in the neighbourhood of the above values of $s_i$, we find that $s_i\approx0.51~\gamma_\bot$ is close to optimal for the case of homodyne detection of the $x$-quadrature. A too small value of $s_i$ (for instance, below $0.25~\gamma_\bot$) leads to increased fluctuations in the predictions of the full model, because the feedback is insufficient to keep the system within the stable region, and a too large value of $s_i$ (for instance, above $1~\gamma_\bot$) tends to course instability. For the case of homodyne detection of the $p$-quadrature, we obtain improved results by decreasing $s_i$ to $0.15~\gamma_\bot$. Trajectories for these parameters are shown in figure~\ref{figure6}. For the upper stable region the standard deviation of the difference between the reduced and the full model relative to the standard deviation of the trajectory obtained from the full model for the time interval from $t=100~\gamma_\bot^{-1}$ to $t=750~\gamma_\bot^{-1}$ is reduced relative to the value obtained for the trajectories in figure~\ref{figure4}. For homodyne detection of the $x$-quadrature it decreases from $0.71$ to $0.46$, and for homodyne detection of the $p$-quadrature it decreases from $1.01$ to $0.82$. This confirms that the integral term feeds back less noise. On the other hand, the integral term is less efficient in keeping $\langle\hat{a}+\hat{a}^\dag\rangle/2$ close to the desired value, and the standard deviation of $\langle\hat{a}+\hat{a}^\dag\rangle/2$ obtained from the full model is observed to increase. The conclusion is the same for the lower stable region for homodyne detection of the $x$-quadrature, but for homodyne detection of the $p$-quadrature we observe a decrease in both the standard deviation of the difference between the reduced and the full model and in the standard deviation of $\langle\hat{a}+\hat{a}^\dag\rangle/2$ obtained from the full model.

\section{Conclusion}\label{6}

In conclusion, we have shown that it is possible to derive simple low-dimensional models for the dynamics of a single atom interacting with a cavity field mode in the absorptive bistable regime, and we have demonstrated that the models can be used to construct a feedback scheme, which is able to hold the system within one of the stable regions. This is an important result, because it is unrealistic to integrate the full stochastic master equation in real time.

The suggested feedback scheme relies on predictions of the expectation value of the $x$-quadrature of the cavity field, and we have considered both the case of homodyne detection of the $x$-quadrature of the output field from the cavity and homodyne detection of the $p$-quadrature. The former case is relatively easy to handle, because the estimated quantity is directly related to the observed quantity. For homodyne detection of the $p$-quadrature, on the other hand, the expectation value of the $x$-quadrature has to be inferred from the precise time evolution of the $p$-quadrature, and the reduced model has to do significantly more work. It is thus promising for the method that we also obtain reasonable results in this case.

There are many degrees of freedom in the modelling procedure and, in general, it may require some trial and error to find reduced models that are able to reproduce the system dynamics with sufficient accuracy. In a feedback scheme, one should concentrate on optimizing the ability of the reduced model to predict the quantity that determines the feedback, since errors in this quantity lead to a feedback of noise into the system, which limits the performance of the feedback.

A further line of research could be to find systematic methods to optimize the models. One could, for instance, consider different ways to construct the vectors used as input to the local tangent space alignment procedure, which corresponds to different criteria for the optimal choice of low-dimensional manifold. It is also possible that improved models could be obtained by choosing other kinds of fitting models than polynomials of low order. The selection of the density operators used to compute the low-dimensional manifold could be adjusted according to the purpose of the model. In the case of application of feedback, one could, for instance, include more points at one attractor than the other in order to obtain a better description of the dynamics in the neighbourhood of the attractor we intend to stabilize.

\ack

A E B Nielsen acknowledges financial support from the Danish Minister of Science, Technology and Innovation through an elite research scholarship.

\end{document}